\documentstyle[aps,prb,epsfig,float,twocolumn]{revtex}
\begin{document}
\twocolumn[\hsize\textwidth\columnwidth\hsize\csname
@twocolumnfalse\endcsname

\title{Variational study of the $\nu=1$ quantum Hall ferromagnet
in the presence of spin-orbit interaction}

\author{John Schliemann, J. Carlos Egues\cite{carlosaddress}, and Daniel Loss}

\address{Department of Physics and Astronomy, University of Basel,
CH-4056 Basel, Switzerland}

\date{\today}

\maketitle

\begin{abstract}
We investigate the $\nu=1$ quantum Hall ferromagnet in the presence of 
spin-orbit coupling of the Rashba or Dresselhaus type by means of
Hartree-Fock-typed variational states. In the presence of
Rashba (Dresselhaus) spin-orbit coupling the fully spin-polarized 
quantum Hall state is {\em always} unstable resulting in a reduction
of the spin polarization if the product of the particle 
charge $q$ and the effective g-factor is positive (negative).
In all other cases an alternative variational state with O(2) symmetry
and finite in-plane spin components 
is lower in energy than the fully spin-polarized state
for large enough spin-orbit interaction. 
The phase diagram resulting from these considerations differs
qualitatively from earlier studies.
\end{abstract}
\vskip2pc]

\section{Introduction}

In the recent years the emerging field of spintronics 
\cite{Wolf01,Awschalom02} has generated an intense interest in effects of
spin-orbit interaction in low-dimensional semiconductor heterostructures.

On the level of effective Hamiltonians arising from $\vec k\cdot\vec p$
theory, the most important effects of spin-orbit interaction 
in low-dimensional geometry are described
by the Rashba term \cite{Rashba60}
and the (linear) Dresselhaus term \cite{Dresselhaus55}. These effective
contributions to the single-particle Hamiltonian stem from the
structure inversion asymmetry of the heterostruture (such as a quantum well)
and the bulk-inversion asymmetry of the semiconductor material, respectively.
In recent years practical manifestations of these kinds of spin-orbit 
interactions have been investigated intensively, see e.g. Refs.
\cite{Pikus88,Gerchikov92,Datta90,Das90,Falko92,Andrada97,Nitta97,Moroz99,Papadakis99,Falko00,Winkler00,Matsuyama00,Grundler00,Grundler01,Hausler01,Mireles01,Aleiner01,Pareek01,Molenkamp01,Nitta02,Matsuyama02,Governale02,Egues02,Miller02}.

Another important topic in low-dimensional semiconductor heterostructures 
is the field of 
quantum Hall ferromagnets \cite{Pinczuk97,Jungwirth01}.
This class of systems includes electron spin ferromagnets as realized
by monolayers at filling factor $\nu=1$, but also bilayer quantum Hall
systems involving a layer (pseudo-)spin \cite{Girvin00}.
Bilayer quantum Hall systems at total filling factor $\nu=1$ have
attracted particular interest very recently due to the spectacular
tunneling experiments by Spielman {\it et al.} \cite{Spielman01}.
In such systems the layer spin is involved in the
most interesting effects such as the spontaneous phase coherence between
the layers, while the electron spin is assumed to be completely aligned 
along the magnetic field and is therefore not of significance.
This is different in bilayer systems at total filling factor $\nu=2$
where both the layer and the electron spin degree of freedom create a
rich phase diagram \cite{Zheng97,MacDonald99,Schliemann00,DasSarma01}.
Yet another type of quantum Hall ferromagnetism occurs in monolayer systems
if single particle states in different Landau level with different 
spins are tuned to energetic coincidence as it can be done by tilting
the magnetic field \cite{Jungwirth98,Jungwirth01}. 
An important connection between electron spin quantum Hall ferromagnetism 
in monolayers and spin-orbit effects arises from 
studies of the dependence of the
effective g-factor on the lattice constant of the semiconductor material 
which can be varied by applying external pressure 
\cite{Usher90,Maude96,Leadley97,Zhitomirsky02}.

In the present work we examine the ground state of a quantum Hall
monolayer at filling factor $\nu=1$ in the presence of spin-orbit coupling
of either the Rashba or the Dresselhaus type. Depending on the type
of spin-orbit interaction, the sign of the charge of the particles
and the sign of their g-factor we find different
kinds of instabilities of the conventional fully spin-polarized
ferromagnetic quantum Hall ground state.
In detail our results are to some extent
in conflict with earlier findings \cite{Falko00}. 

The paper is organized as follows. In section \ref{single-particle} we
review the single-particle states of charged particles in two-dimensional
layers in the presence of a perpendicular magnetic field and spin-orbit 
interaction of the aforementioned type \cite{Das90,Falko92}. In section
\ref{variational} we present our variational approach with several technical 
details given in the two appendices. We close with a discussion of our 
results in section \ref{discussion}.


\section{Single-particle states}
\label{single-particle}

We consider a spin-$\frac{1}{2}$-particle of charge \cite{note1}
$q=\mp|e|$ and effective mass $m$ moving in a two-dimensional ($xy$)-plane 
provided by a semiconductor quantum well. The particle is subject to
spin-orbit interaction and to a perpendicular magnetic field 
$\vec B=B\vec e_{z}=\nabla\times\vec A$ which couples to the orbital and
the spin degree of freedom. The Hamiltonian reads using standard notation
\begin{equation}
{\cal H}=\frac{1}{2m}\left(\vec p-\frac{q}{c}\vec A\right)^{2}
+\frac{1}{2}g\mu_{B}B\sigma^{z}+{\cal H}_{so}
\end{equation}
where $g$ is the effective g-factor of the particle and
$\mu_{B}=|e|\hbar/(2m_{0}c)$ the Bohr magneton with $m_{0}$ being the
bare electron mass. 
In a semiconductor heterostructure such as a 
quantum well the spin-orbit coupling in the conduction band 
has, for appropriate growth geometry,
two relevant contributions, ${\cal H}_{so}={\cal H}_{R}+{\cal H}_{D}$ with
\begin{eqnarray}
{\cal H}_{R} & = & \alpha\left(\pi_{x}\sigma^{y}-\pi_{y}\sigma^{x}\right)\\
{\cal H}_{D} & = & \beta\left(\pi_{x}\sigma^{x}-\pi_{y}\sigma^{y}\right)
\label{soham}
\end{eqnarray}
where we have introduced the kinetic momentum 
$\vec \pi=\vec p-\frac{q}{c}\vec A$. The first term ${\cal H}_{R}$ is the
Rashba spin-orbit coupling arising from the structure inversion asymmetry
of the quantum well, and the second contribution is the (linear) Dresselhaus
term which stems from the bulk-inversion asymmetry of the semiconductor 
material. The coefficient $\beta$ of the Dresselhaus term is fully determined
by the geometry of the hetereostructure while the Rashba coefficient
$\alpha$ can be varied by an electric field across the well \cite{Nitta97}.
We note that the Rashba Hamiltonian has an SU(2) symmetry (under simultaneous
rotations of kinetic momentum and spin), while the symmetry group of
the Dresselhaus term is SU(1,1). 

Defining the usual bosonic operators
\begin{equation}
a=\frac{1}{\sqrt{2}}\frac{\ell}{\hbar}\left(\pi_{x}+i\delta\pi_{y}\right)
\quad,\quad a^{+}=(a)^{+}
\end{equation}
with $[a,a^{+}]=1$, $\delta={\rm sgn}(qB)$ and $\ell=\sqrt{\hbar c/|qB|}$
being the magnetic length, the Hamiltonian reads
\begin{equation}
{\cal H}=\hbar\omega_{c}\left(a^{+}a+\frac{1}{2}\right)
+\frac{1}{2}g\mu_{B}B\sigma^{z}+{\cal H}_{so}
\end{equation}
Here $\omega_{c}=|qB|/(mc)$ is the cyclotron frequency, and the spin-orbit
contributions take the form
\begin{eqnarray}
{\cal H}_{R} & = & \left\{
\begin{array}{cc}
\frac{i}{\sqrt{2}}
\alpha\frac{\hbar}{\ell}\left(a\sigma^{-}-a^{+}\sigma^{+}\right)
& \quad\delta=+1\\ & \\
\frac{i}{\sqrt{2}}
\alpha\frac{\hbar}{\ell}\left(a^{+}\sigma^{-}-a\sigma^{+}\right)
& \quad\delta=-1
\end{array}\right.
\\
{\cal H}_{D} & = & \left\{
\begin{array}{cc}
\frac{1}{\sqrt{2}}
\beta\frac{\hbar}{\ell}\left(a\sigma^{+}+a^{+}\sigma^{-}\right)
& \quad\delta=+1\\ & \\
\frac{1}{\sqrt{2}}
\beta\frac{\hbar}{\ell}\left(a\sigma^{-}+a^{+}\sigma^{+}\right)
& \quad\delta=-1
\end{array}\right.
\end{eqnarray}
with $\sigma^{\pm}=\sigma^{x}\pm i\sigma^{y}$.  
The operators $a$ and $a^{+}$ connect different Landau levels. Another
set of important operators is given in terms of the components of the
center $\vec r_{0}=(x_{0},y_{0})$ of the classical orbital motion and read
\begin{equation}
b=\frac{1}{\sqrt{2}}\frac{1}{\ell}\left(x_{0}-i\delta y_{0}\right)
\quad,\quad b^{+}=(b)^{+}
\end{equation}
These operators fulfil $[b,b^{+}]=1$, commute with $a$, $a^{+}$ and connect
different orbital states within a given Landau level. Since the Hamiltonian
(including the spin-orbit part) can be expressed in terms of $a$ and $a^{+}$
only, its eigenstates have the same Landau level degeneracy as in the absence
of spin-orbit coupling.

In the presence of both Rashba and the Dresselhaus term,
the spin-orbit interaction couples all states in all Landau levels, and 
an analytical solution to the full problem is unknown
\cite{Das90,Falko92}. Therefore we shall
restrict ourselves to the case where either only Rashba or Dresselhaus 
spin-orbit coupling is present, and the Hamiltonian commutes with the 
operator $L=a^{+}a\mp\delta\sigma^{z}/2$ which can be used to classify
eigenstates.
Fixing a certain intra-Landau-level
quantum number, we denote by 
$|n,\sigma\rangle=((a^{+})^{n}/\sqrt{n!})|0,\sigma\rangle$ 
a state in the $n$-th
Landau level with spin direction $\sigma\in\{\uparrow,\downarrow\}$.
Without loss of generality we discuss the case of Rashba 
coupling with $\delta>0$. Then $|0,\uparrow\rangle$ is an eigenstate
with energy $\varepsilon_{0}=(\hbar\omega_{c}+g\mu_{B}B)/2$ and $L=-1/2$.
All other eigenstates are of the form \cite{Das90,Falko92}
\begin{equation}
|n,\pm\rangle=u^{\pm}_{n}|n,\uparrow\rangle+v^{\pm}_{n}|n-1,\downarrow\rangle
\label{speigen}
\end{equation}
with $L=n-1/2$, $n>0$, and energy
\begin{equation}
\varepsilon_{n}^{\pm}=\hbar\omega_{c}n\pm
\sqrt{2n\alpha^{2}m\hbar\omega_{c}+\frac{1}{4}
\left(\hbar\omega_{c}+g\mu_{B}B\right)^{2}}
\end{equation}
and the amplitudes
parametrizing the eigenstates read
\begin{eqnarray}
u^{\pm}_{n} & = & 
\left(\frac{1}{2}\pm\frac{\frac{1}{4}\left(\hbar\omega_{c}+g\mu_{B}B\right)}
{\sqrt{2n\alpha^{2}m\hbar\omega_{c}+\frac{1}{4}
\left(\hbar\omega_{c}+g\mu_{B}B\right)^{2}}}\right)^{\frac{1}{2}}
\label{amp1}\\
v^{\pm}_{n} & = & \pm i\,{\rm sgn}(\alpha)\nonumber\\
 & & \cdot\left(\frac{1}{2}\mp
\frac{\frac{1}{4}\left(\hbar\omega_{c}+g\mu_{B}B\right)}
{\sqrt{2n\alpha^{2}m\hbar\omega_{c}+\frac{1}{4}
\left(\hbar\omega_{c}+g\mu_{B}B\right)^{2}}}\right)^{\frac{1}{2}}
\label{amp2}
\end{eqnarray}
The single-particle eigenstates for the case $\delta<0$ and/or
Dresselhaus instead of Rashba coupling can be obtained 
by obvious modifications of the above expressions; in figure
\ref{fig1} we give a schematic overview of the coupling of Landau levels
due to the two different types of spin-orbit interaction.
Note that the above solution does not require the specification of a gauge 
for the vector potential creating the magnetic field.

The lowest single-particle states are given by $|0,\uparrow\rangle$ and 
$|1,-\rangle$ which will be of particular interest in the following. 
The latter state is lower in energy than the first one, i.e. 
$\varepsilon_{1}^{-}-\varepsilon_{0}<0$, if 
\begin{equation}
-g\mu_{B}B<2\alpha^{2}m
\label{cond}
\end{equation}
This condition involves the Zeeman energy $\Delta_{z}=-g\mu_{B}B$
and the Rashba energy scale $\alpha^{2}m$, but remarkably not the
cyclotron energy $\hbar\omega_{c}$. Provided that $\hbar\omega_{c}>g\mu_{B}B$
the above condition is not only a sufficient but also a necessary
criterion for $\varepsilon_{1}^{-}-\varepsilon_{0}<0$.
The inequality (\ref{cond})
will be of crucial importance in the following section.


\section{Variational approach to the $\nu=1$ quantum Hall ferromagnet}
\label{variational}

We now investigate variational {\em ans\"atze} for the ground state of
the two-dimensional electron gas at filling factor $\nu=1$ in the presence
of Coulomb interaction and spin-orbit coupling. Assuming the ground state
to be spatially homogeneous, we consider as variational states Slater
determinants consisting of single particle states having each a different 
intra-Landau-level quantum number, and all these quantum numbers are covered 
leading to a filling factor of unity.

\subsection{Rashba coupling with $gq<0$, or Dresselhaus coupling
with $gq>0$}
\label{case1}

Let us first consider variational states appropriate for particles being 
subject to Rashba spin-orbit interaction and having charge $q$ and g-factor 
$g$ fulfilling $qg<0$, or, alternatively, Dresselhaus coupling with
$qg>0$. To be specific we investigate electrons ($q=-|e|<0$) with
positive g-factor in the presence of Rashba coupling. All other cases
can be derived from this one by obvious modifications.

Without loss of generality we choose the magnetic field to point in the
negative $z$-direction, $\vec B=-|B|\vec e_{z}$, i.e. $\delta>0$.
Since the g-factor is positive the Zeeman term favors the electron spin to 
align antiparallel to the magnetic field, i.e. in the positive
$z$-direction. Therefore we consider as a variational {\em ansatz} 
for the ground state the Slater determinant constructed from all 
single-particle states of the form (cf. Eq.~(\ref{speigen})),
\begin{equation}
p|0,\uparrow,m\rangle+r|1,-,m\rangle
\label{ansatz1}
\end{equation}
where the variational parameters $p$ and $r$ are subject to the normalization 
condition $|p|^{2}+|r|^{2}=1$. $m$ is some intra-Landau-level index
parametrizing all degenerate single-particle states of the above form.
Thus our variational state is a Slater determinant built up from
all these single-particle states yielding a filling factor of $\nu=1$.

We now study the energy of our variational state in the presence of
Coulomb interaction including a neutralizing background.
With the results of appendix \ref{coulombapp}
 one finds for the energy per particle
\begin{eqnarray}
\varepsilon^{(1)}_{var} & = & \varepsilon_{fp}^{\uparrow}
+\left(\varepsilon_{1}^{-}-\varepsilon_{0}\right)|r|^{2}\nonumber\\
& + & \left[\frac{e^{2}}{\kappa\ell}\sqrt{\frac{\pi}{8}}
\frac{1}{2}|u_{1}^{-}|^{2}
\left(\frac{1}{2}|u_{1}^{-}|^{2}+2|v_{1}^{-}|^{2}\right)\right]|r|^{4}
\label{varerg1}
\end{eqnarray}
where $\kappa$ is the dielectric constant of the semiconductor material
and the coeffcients $u_{1}^{-}$ and $v_{1}^{-}$ are given by 
Eqs.~(\ref{amp1}),(\ref{amp2}).
\begin{equation}
\varepsilon_{fp}^{\uparrow}=\frac{1}{2}
\left(\hbar\omega_{c}+g\mu_{B}B\right)
-\frac{e^{2}}{\kappa\ell}\sqrt{\frac{\pi}{8}}
\label{ferroerg}
\end{equation}
is the energy per particle of the
variational state at $r=0$, i.e. the
$\nu=1$ ferromagnetic quantum Hall state lying purely in the lowest
Landau level
with all spins pointing along the positive $z$-axis (which is the preferred
direction for $gB<0$.)

The variational energy (\ref{varerg1}) becomes smaller than
$\varepsilon_{fp}^{\uparrow}$ for certain values of $|r|$ if and only if
the coefficient $(\varepsilon_{1}^{-}-\varepsilon_{0})$ of the
quadratic term is negative which is equivalent to the condition 
(\ref{cond}). In this case the minimizing value for $|r|$ is given by
\begin{equation}
|r|^{2}=\min\left\{1,\frac{|\varepsilon_{1}^{-}-\varepsilon_{0}|}
{\frac{e^{2}}{\kappa\ell}\sqrt{\frac{\pi}{8}}|u_{1}^{-}|^{2}
\left(\frac{1}{2}|u_{1}^{-}|^{2}+2|v_{1}^{-}|^{2}\right)}\right\}
\end{equation}
with a variational ground state energy of (for $|r|<1$)
\begin{equation}
\varepsilon^{(1)}_{var}=\varepsilon_{fp}^{\uparrow}
-\frac{1}{2}\frac{|\varepsilon_{1}^{-}-\varepsilon_{0}|^{2}}
{\frac{e^{2}}{\kappa\ell}\sqrt{\frac{\pi}{8}}|u_{1}^{-}|^{2}
\left(\frac{1}{2}|u_{1}^{-}|^{2}+2|v_{1}^{-}|^{2}\right)}
\end{equation}
Thus we have found a variational state being lower in energy than
the conventional quantum Hall ferromagnet (having an energy per particle
of $\varepsilon_{fp}^{\uparrow}$).

The variational ground state has a uniform spin density with 
a reduced $z$-component and non-zero in-plane components.
The spin expectation values per
particle read
\begin{eqnarray} 
\langle s^{z}\rangle  & = & \frac{\hbar}{2}
\left(\left(1-|r|^{2}\right)+
|r|^{2}\left(|u^{-}_{1}|^{2}-|v^{-}_{1}|^{2}\right)\right)\\
\langle s^{+}\rangle & = & \hbar p^{*}rv_{1}^{-}
\label{inplane}
\end{eqnarray}
The expectation values of the in-plane components
depend on the relative phase between $p$ and $r$ which does not
influence the energy. Thus the variational ground state reflects
an O(2) symmetry. A smilar instability of the conventional
ferromagnetic quantum Hall state (for pure Rashba coupling)
was also found recently by
Falko and Iordanskii \cite{Falko00} who studied perturbative
expansions of the thermodynamic potential in a path integral formulation.
However, these authors find a different condition for the instability
of the conventional ferromagnetic state which reads in the notation used
here
\begin{equation}
\left|-g\mu_{B}B+\alpha^{2}m\right|
<2\alpha^{2}m
\frac{\frac{e^{2}}{\kappa\ell}\sqrt{\frac{\pi}{8}}}{\hbar\omega_{c}}
\label{Falkocond}
\end{equation}
This differs, especially 
due to the magnetic field dependence of the right hand side,
qualitatively and quantitatively
from our result (\ref{cond}). In particular, for appropriate
parameters the results of Ref.~\cite{Falko00} predict the stability
of the conventional ferromagnetic state while our variational approach
rigorously establishes the existence of a state lower in energy.

\begin{figure}
\centerline{\includegraphics[width=8cm]{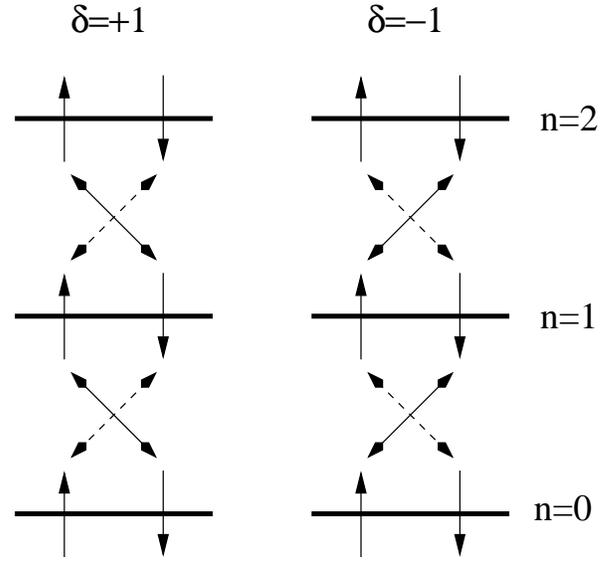}} 
\caption{Schematic of the spin-orbit interaction coupling
Landau levels with different spin directions (symbolized by single arrows).
The solid (dashed) double-arrow lines connect states coupled
by the Rashba (Dresselhaus) spin-orbit interaction.
\label{fig1}}
\end{figure}

Quantum wells with a positive electron g-factor can be realized in terms of
biased GaAs/AlGaAs strcutures \cite{Ivchenko97,Jiang01,Salis01}.
In figure~\ref{fig1} we have plotted the difference $\delta\varepsilon^{(1)}$ 
of the minimum variational energy
$\varepsilon^{(1)}_{var}$ per particle and $\varepsilon_{fp}^{\uparrow}$,
as a function of $|B|$ for electrons with an effective mass of $0.2$ times
the bare electron mass $m_{0}$, 
$g=1$, Rashba energy $\alpha^{2}m=0.5{\rm meV}$,
and a dielectric constant of $\kappa=10.0$. For this choice of parameters 
this quantity is, for magnetic fields $|B|$ between 2 and 5 Tesla, 
of order $0.5{\rm K}$ which should be resolvable by experimental cooling
techniques. 

\begin{figure}
\centerline{\includegraphics[width=8cm]{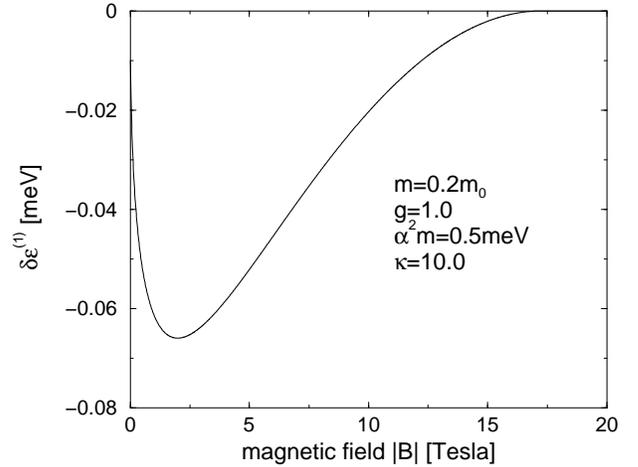}} 
\caption{The difference $\delta\varepsilon^{(1)}$ 
of the minimum variational energy
$\varepsilon^{(1)}_{var}$ per particle and $\varepsilon_{fp}^{\uparrow}$,
as a function of $|B|$ for electrons with an effective mass of $0.2$ times
the bare electron mass $m_{0}$, 
$g=1$, Rashba energy $\alpha^{2}m=0.5{\rm meV}$,
and a dielectric constant of $\kappa=10.0$.
\label{fig2}}
\end{figure}

Figure~\ref{fig2} shows spin expectation values as a function of $|B|$ for
the same system parameters as in figure ~\ref{fig1}. As seen, the in-plane
spin components can be of substantial magnitude.

The variational {\em ans\"atze} studied so far are Slater determinants
consisting of single-particle states being linear combinations
of two different states: One of the states lies completely 
in the lowest Landau level and is not perturbed by the spin-orbit coupling
(in the above example $|0,\uparrow,m\rangle$)
and another state closest in energy which is modified by the spin-orbit
interaction and has a contribution from the first excited Landau level
(in the above example $||1,-,m\rangle$). The spin direction of the
unperturbed state involved is determined by $\delta={\rm sgn}(qB)$ and the
type of spin-orbit coupling. In all combinations
investigated here of the sign of $gq$ and the type of spin-orbit coupling,
the spin of the unperturbed state points into the direction
favored by the Zeeman coupling. Therefore our {\em ansatz} does not 
frustrate the Zeeman coupling. In the following subsection we will
study the opposite cases.

\subsection{Rashba coupling with $gq>0$, or Dresselhaus coupling
with $gq<0$}
\label{case2}

We now investigate the case of Rashba spin-orbit coupling with $gq>0$ or,
alternatively, Dresselhaus coupling with $gq<0$. In these cases the 
variational {\em ansatz} studied in the previous subsection
frustrates the Zeeman coupling, and an alternative variational
state (possibly competing with the first one) becomes appropriate.

As before and  without loss of generality we concentrate on Rashba coupling
of electrons ($q<0$) in a magnetic field pointing in the negative
$z$-direction ($\delta>0$). Since the g-factor is by assumption negative
($gq>0$) our variational {\em ansatz} is a Slater determinant built up from
single-particle states of the form
\begin{equation}
s|0,\downarrow,m\rangle+t|1,\uparrow,m\rangle
\label{ansatz2}
\end{equation}
with variational parameters $s$ and $r$ restricted by
$|s|^{2}+|t|^{2}=1$, and $m$ is again some intra-Landau-level index.
Choosing $s^{*}t=i{\rm sgn}(\alpha)|s^{*}t|$ the variational ground state
energy per particle is given by
\begin{eqnarray}
\varepsilon^{(2)}_{var} & = & \varepsilon_{fp}^{\downarrow}
+\left(\hbar\omega_{c}+g\mu_{B}B\right)|t|^{2}
-\frac{4|\alpha|\hbar}{\sqrt{2}\ell}|t|\sqrt{1-|t|^{2}}
\nonumber\\
& + & \frac{e^{2}}{\kappa\ell}\sqrt{\frac{\pi}{8}}
|t|^{2}\left(1-\frac{3}{4}|t|^{2}\right)
\label{varerg2}\\
 & = & \varepsilon_{fp}^{\downarrow}-\frac{4|\alpha|\hbar}{\sqrt{2}\ell}|t|
\nonumber\\
& + &\left(\hbar\omega_{c}+g\mu_{B}B
+\frac{e^{2}}{\kappa\ell}\sqrt{\frac{\pi}{8}}\right)|t|^{2}
+{\cal O}\left(|t|^{3}\right)
\end{eqnarray}
with $\varepsilon_{fp}^{\downarrow}=\varepsilon_{fp}^{\uparrow}-g\mu_{B}B$
being the energy per particle of the conventional ferromagnetic state
in the absence of spin-orbit coupling. Since the coefficient of the term linear
in $|t|$ is negative (for the above choice of phases) the minimum
$\varepsilon_{var}^{(2)}$ 
is always smaller than $\varepsilon_{fp}^{\downarrow}$.
Note that this observation is independent of the sign of the g-factor and
occurs also for the cases studied in the previous subsection. This shows that
the fully spin-polarized quantum Hall state in the lowest Landau level
is strictly speaking {\em always} unstable in the presence 
of spin-orbit coupling. This holds also if both Rashba and Dresselhaus
coupling are present. Within the {\em ansatz} (\ref{ansatz2}) for instance
the Dresselhaus term does not contribute to the energy expectation
value at all. Therefore this variational {\em ansatz} might not be
optimal for this more general case but still yields a variational energy
lower than  $\varepsilon_{fp}^{\downarrow}$. However, in the cases 
investigated in the previous subsection \ref{case1} the 
{\em ansatz} (\ref{ansatz1}) usually gives lower energies than 
(\ref{ansatz2}) for realistic system parameters fulfilling the inequality
(\ref{cond}). In the remainder of this subsection we shall concentrate again
on the case of Rashba spin-orbit coupling with $gq>0$.

In the variational state (\ref{ansatz2})
the $z$-component of the spin per particle is
reduced and given by
\begin{equation}
\langle s^{z}\rangle=-\frac{\hbar}{2}\left(1-2|t|^{2}\right)
\label{sz2}
\end{equation}
The in-plane spin components identically vanish 
within the above variational state. This might appear as an artifact of
the {\em ansatz} used here and could be altered if the other spin direction
in the lowest Landau level is also taken into account.
Such a generalized variational state is studied in appendix \ref{genapp}.
It turns out that this generalized {\em ansatz} does not lead
to variational energies lower than obtained so far if the cyclotron
energy $\hbar\omega$ is larger in magnitude than the Zeeman splitting   
$\Delta_{z}=-g\mu_{B}B$. This is usually the case in semiconductors.
Therefore, within our variational approach, the only effect of
spin orbit coupling under the conditions discussed in this subsection
is to reduce the magnetization of the quantum Hall ground state but not
to alter its direction. The case considered here (electrons with negative
g-factor and Rashba coupling) includes the important case of conduction band
electrons in the III-V
semiconductors GaAs, InAs, and InSb. For InAs a typical value for the 
Rashba energy $\alpha^{2}m$ is $0.5{\rm meV}$ \cite{Grundler00}. 
With this number and the
material parameters for InAs we find the reduction of the magnetization 
according to Eq.~(\ref{sz2}) (for the minimizing value of $t$) to be
a few percent at typical fields $|B|$ of a few Tesla. However, the 
reduction becomes more pronounced with decreasing modulus of the g-factor.


\section{Discussion}
\label{discussion}

We have studied the effect of spin-orbit coupling on the ground state
of a $\nu=1$ quantum Hall monolayer using variational Hartree-Fock-typed
states, which are Slater determinants consisting of linear combinations 
of low-lying single-particle states. 

In the case of Rashba coupling and the product of the charge $q$ and the 
effective g-factor $g$ of the particles being negative we find an instability
of the spin-polarized ferromagnetic state toward a state with O(2)
symmetry. The same result is valid for the formally equivalent case of
Dresselhaus coupling with $gq>0$. These results are the similar to the recent
findings by Falko and Iordanskii \cite{Falko00}
(for pure Rashba or Dresselhaus coupling), although we obtain a
qualitatively and quantitatively 
different phase boundary between both types of states.

For the opposite cases (Rashba coupling with $gq>0$, or Dresselhaus coupling
with $gq<0$) we have used a variational {\em ansatz} which involves
all relevant low-lying single-particle states: the two states in the
lowest Landau level for the two spin directions, and the states in the
first excited Landau level with the the appropriate spin direction
such that this state is coupled to the lowest Landau level by the spin-orbit
interaction. We therefore believe that this variational state captures
the essential ground state properties.
As a result,  we do not find an instability toward an O(2) symmetric
ground state with finite in-plane spin components, 
but just a (typically small) reduction of the magnetization
with its direction being unaltered. Moreover, this instability always occurs
and does not depend on other system parameters. 
These results are further important differences
from the 
findings of Ref.~\cite{Falko00}, where, depending on system parameters,
a deviation of the
magnetization from the field direction in the ground state was predicted.
The reduction of the magnetization increases with decreasing modulus of the
Zeeman splitting. Therefore spin-orbit effects can be a part of the
explanation for recent experimental data by Zhitomirsky {\it et al.}
\cite{Zhitomirsky02}, where a quantum Hall state at $\nu=1$ with incomplete
spin polarization was reported for small g-factors.

The Rashba coupling with  $gq>0$ covers in particular the
important case of conduction band electrons in III-V semiconductors such as 
GaAs, InAs, and InSb. 
Thus, our results indicating only a small reduction
in magnetization can be seen as good news with respect to 
proposals to use integer 
quantum Hall systems as sources of spin-polarized electrons in 
experiments related to spintronics and quantum information processing
\cite{Recher00,Vandersypen02}.

It is interesting to speculate how spin-orbit interactions might affect other
types of quantum Hall ferromagnets \cite{Pinczuk97}. In the important case of
bilayers at total filling factor $\nu=1$ the
electron spins are assumed to be polarized by the magnetic field, while
the layer pseudospin forms an easy-plane typed
ferromagnetic ground state showing a very robust spontaneous symmetry breaking
\cite{Spielman01}. In this case we do not expect the spin-orbit coupling
to the electron spin to contribute substantially to those physical properties.
The situation is different for bilayer systems at total filling factor
$\nu=2$, where both
the electron and the layer spin are involved in various phase transitions
\cite{Zheng97,MacDonald99,Schliemann00,DasSarma01}.
In this case we expect the spin-orbit interaction to even enrich
the phase diagram of the system. Similarly, spin-orbit coupling might
also alter the phase transitions predicted in monolayers when two different
Landau level with different spin directions are tuned to energetic
coincidence \cite{Jungwirth98,Jungwirth01}.


\acknowledgements{This work has been supported by the Swiss NSF, 
NCCR Nanoscience, DARPA, and ARO.}
 

\appendix


\section{The Coulomb energy of the variational states}
\label{coulombapp}

The Coulomb energy of the variational states investigated in this paper
can be obtained via evaluating the pair distribution function for
many-body Slater determinants $|\Psi\rangle$ constructed from all
possible single-particle states of the form
\begin{equation}
a|0,\uparrow,m\rangle+b|0,\downarrow,m\rangle+c|1,\uparrow,m\rangle
\label{gensp}
\end{equation}
with $|a|^{2}+|b|^{2}+|c|^{2}=1$.  
Here $m$ is some intra-Landau-level index, and 
$ |1,\uparrow,m\rangle=a^{+}|0,\uparrow,m\rangle$.
$|\Psi\rangle$ contains all
single-particle states of this type leading to a filling factor of unity.
To compute the pair distribution function
\begin{equation}
g(\vec r_{1}-\vec r_{2})=
\langle\Psi|\sum_{i\neq j}
\left[\delta
(\vec r_{1}-\hat{\vec r_{i}})\delta(\vec r_{2}-\hat{\vec r_{j}})\right]
|\Psi\rangle
\end{equation}
it is convenient to work in the symmetric gauge $\vec A=B(-y,x,0)/2$
assuming (without loss of generality) $\delta>0$
with orbital part of the wave functions given by (supressing the spin index)
\begin{eqnarray}
\psi_{0,m}(z) & := & \langle\vec r|0,m\rangle \nonumber\\
& = & \frac{1}{\sqrt{2\pi\ell^{2}m!}}\left(\frac{z}{\sqrt{2}\ell}\right)^{m}
\exp\left(-\frac{|z|^{2}}{4\ell^{2}}\right)\\
\psi_{1,m-1}(z) & := & \langle\vec r|1,m\rangle
=\frac{-i}{\sqrt{2\pi\ell^{2}m!}}\left(\frac{z}{\sqrt{2}\ell}\right)^{m-1}
\nonumber\\
&  & \cdot\left(m-\frac{|z|^{2}}{2\ell^{2}}\right)
\exp\left(-\frac{|z|^{2}}{4\ell^{2}}\right)
\end{eqnarray}
where $z=x+iy$ and the second subscript in the wave functions denotes
the eigenvalue of the angular momentum 
$M=\delta\hbar(b^{+}b-a^{+}a)$.

A straightforward calculation leads to the following expression for
the pair distribution function
\begin{eqnarray}
g(\vec r) & = & \frac{1}{(2\pi\ell^{2})^{2}}\Biggl[1-\Biggr(1+
|c|^{4}\left(\left(1-\frac{r^{2}}{2\ell^{2}}\right)^{2}-1\right)\nonumber\\
 & & \qquad
-2|b|^{2}|c|^{2}\left(1-\frac{r^{2}}{2\ell^{2}}\right)\nonumber\\
 & & \qquad
+2\Re\left\{a^{2}c^{*2}\frac{x^{2}-y^{2}+2ixy}{2\ell^{2}}\right\}
\Biggr)\nonumber\\
 & & \quad\cdot\exp\left(-\frac{r^{2}}{2\ell^{2}}\right)\Biggr]
\label{pdf}
\end{eqnarray}
To obtain this result we have used the relations
\begin{eqnarray}
& & \sum_{m=0}^{\infty}\psi_{0,m}^{*}(z_{1})\psi_{0,m}(z_{2})
=\frac{1}{2\pi\ell^{2}}\nonumber\\
& & 
\qquad\cdot\exp\left(\frac{z_{1}^{*}z_{2}}{2\ell^{2}}
-\frac{|z_{1}|^{2}+|z_{2}|^{2}}{4\ell^{2}}\right)\\
& & \sum_{m=0}^{\infty}\psi_{0,m}^{*}(z_{1})\psi_{1,m-1}(z_{2})
=\frac{-i}{2\pi\ell^{2}}\frac{z_{1}^{*}-z_{2}^{*}}{\sqrt{2}\ell}\nonumber\\
& & 
\qquad\cdot\exp\left(\frac{z_{1}^{*}z_{2}}{2\ell^{2}}
-\frac{|z_{1}|^{2}+|z_{2}|^{2}}{4\ell^{2}}\right)\\
& & \sum_{m=0}^{\infty}\psi_{1,m-1}^{*}(z_{1})\psi_{1,m-1}(z_{2})
=\frac{1}{2\pi\ell^{2}}
\left(1-\frac{|z_{1}-z_{2}|^{2}}{2\ell^{2}}\right)\nonumber\\
& & 
\qquad\cdot\exp\left(\frac{z_{1}^{*}z_{2}}{2\ell^{2}}
-\frac{|z_{1}|^{2}+|z_{2}|^{2}}{4\ell^{2}}\right)
\end{eqnarray}
The first term in the rectangular brackets in Eq.~(\ref{pdf}) is the
Hartree contribution to the pair distribution function, while the
expression proportional to the exponential is the Fock term.
Note that for $c=0$ the state $|\Psi\rangle$ is just the
usual $\nu=1$ ferromagnet in the lowest Landau level with its sponteneous
spin polarization parametrized by the coefficients $b$ and $c$, and
the pair distribution function reduces 
to its well-known expression for this case.
  
The pair distribution function $g(\vec r)$ contains  non-isotropic
contributions (i.e. terms are not functions of $r=\sqrt{x^{2}+y^{2}}$ only).
This is due to the fact that the single-particle states (\ref{gensp})
involve (for $a\neq0\neq c$) superpositions of states with the same spin but
different orbital angular momentum. However, these non-isotropic terms
do not contribute to integrals of the form
$\int d^{2}rf(r)g(\vec r)$. In particular, for the Coulomb interaction energy
per particle in the presence of a neutralizing background one finds
\begin{equation}
\varepsilon_{c}=-\frac{e^{2}}{\kappa\ell}\sqrt{\frac{\pi}{8}}
\left(1-\frac{1}{2}|c|^{2}\left(\frac{1}{2}|c|^{2}+2|b|^{2}\right)\right)
\end{equation}


\section{The generalized variational ansatz}
\label{genapp}

In this appendix we discuss a generalized variational {\em ansatz} where
the single-particle states are arbitrary linear combinations (for a given
intra-Landau level quantum number) of both states in the lowest Landau level
(for both spin directions) and the state in the first excited Landau level
with the appropriate spin direction coupled by the spin-orbit interaction
to the lowest Landau level. We again concentrate
on the case of Rashba coupling of electrons ($q=-|e|<0$) in a magnetic
field $\vec B=-|B|\vec e_{z}$, i.e. $\delta>0$.
In this case our variational many-body state is a Slater determinant 
consisting of single particle states of the form (cf. Eq.~(\ref{gensp}))
\begin{equation}
a|0,\uparrow,m\rangle+b|0,\downarrow,m\rangle+c|1,\uparrow,m\rangle
\end{equation}
with $|a|^{2}+|b|^{2}+|c|^{2}=1$. 
Choosing $b^{*}c=i{\rm sgn}(\alpha)|b^{*}c|$ the variational ground state
energy per particle is given by
\begin{eqnarray}
\varepsilon_{var} & = & \varepsilon_{fp}^{\uparrow}
+\Delta_{z}|b|^{2}+\hbar\omega_{c}|c|^{2}
-\frac{4|\alpha|\hbar}{\sqrt{2}\ell}|b||c|\nonumber\\
& + & \frac{e^{2}}{\kappa\ell}\sqrt{\frac{\pi}{8}}
\frac{1}{2}|c|^{2}\left(\frac{1}{2}|c|^{2}+2|b|^{2}\right)
\end{eqnarray}
where $\varepsilon_{fp}^{\uparrow}$ is given by Eq.~(\ref{ferroerg})
and we have introduced the Zeeman splitting
$\Delta_{z}=-g\mu_{B}B=g\mu_{B}|B|$. We now search for the minimum of the
above variational energy under the normalization restriction 
$|b|^{2}+|c|^{2}\leq 1$. This minimum can either lie on the edge of the
allowed range ($b|^{2}+|c|^{2}=1$) or in its interior. In the latter case
the stationarity condition 
$(\partial\varepsilon_{var}/\partial|b|)=
(\partial\varepsilon_{var}/\partial|c|)=0$ holds from which one finds the 
relations
\begin{equation}
|b|=\frac{\frac{4|\alpha|\hbar}{\sqrt{2}\ell}|c|}
{2\left(\Delta_{z}
+\frac{e^{2}}{\kappa\ell}\sqrt{\frac{\pi}{8}}|c|^{2}\right)}
\label{varrel1}
\end{equation}
\begin{equation}
2\Delta_{z}|b|^{2}-\hbar\omega_{c}|c|^{2}-
\frac{e^{2}}{\kappa\ell}\sqrt{\frac{\pi}{8}}|c|^{4}=0
\label{varrel2}
\end{equation}
These equations 
have the obvious solution $|b|=|c|=0$ corresponding to a conventional
ferromagnetic quantum Hall state. As seen in subsection \ref{case2} this
is not an energy minimum.
For $|b|\neq 0\neq |c|$ inserting 
(\ref{varrel1}) into (\ref{varrel2}) yields a third-order polynomial
equation for $x:=|c|^{2}$,
\begin{equation}
x^{3}+rx^{2}+sx+t=0
\label{poly1}
\end{equation}
with
\begin{eqnarray}
r & = & \frac{2\left(\hbar\omega_{c}+\Delta_{z}\right)}
{\frac{e^{2}}{\kappa\ell}\sqrt{\frac{\pi}{8}}}\\
s & = & \frac{\Delta_{z}\left(4\hbar\omega_{c}+\Delta_{z}\right)}
{\left(\frac{e^{2}}{\kappa\ell}\sqrt{\frac{\pi}{8}}\right)^{2}}\\
t & = & \frac{2\hbar\omega_{c}\Delta_{z}
\left(\Delta_{z}-2\alpha^{2}m\right)}
{\left(\frac{e^{2}}{\kappa\ell}\sqrt{\frac{\pi}{8}}\right)^{3}}
\end{eqnarray}
For $\Delta_{z}>0$ (corresponding to $g>0$, cf. subsection \ref{case1})
$r$ and $s$ are both positive. Therefore the above equation can only have a
positive solution if $\Delta_{z}-2\alpha^{2}m<0$, in accordance with
our criterion (\ref{cond}) for the instability of the conventional
ferromagnetic state. The latter result was obtained in subsection \ref{case1}
within a restricted
{\em ansatz} where the ratio of $c/b$ is fixed according to the 
single-particle amplitudes (\ref{amp1}), (\ref{amp2}).
Our result here shows that the generalized variational state (\ref{gensp})
leads to the same stability region, i.e. 
qualitatively to the identical phase diagram. The more general {\em ansatz}
(\ref{gensp}) might in principle give even lower 
energy minima than obtained before. However, this cannot alter the
qualitative results and we shall not further discuss this issue here.  

Let us now turn to the case 
$\Delta_{z}<0$ (corresponding to $g<0$, cf. subsection \ref{case2}).
With the standard substitution $x=:y-r/3$ one finds
\begin{equation}
y^{3}+py+q=0
\label{poly2}
\end{equation}
with 
\begin{eqnarray}
p & = & 
\frac{-1}{\left(\frac{e^{2}}{\kappa\ell}\sqrt{\frac{\pi}{8}}\right)^{2}}
\frac{4}{3}\left(\hbar\omega_{c}-\frac{1}{2}\Delta_{z}\right)^{2}\leq 0\\
q & = &
\frac{1}{\left(\frac{e^{2}}{\kappa\ell}\sqrt{\frac{\pi}{8}}\right)^{3}}
\nonumber\\
& & \cdot
\frac{16}{27}\left(\left(\hbar\omega_{c}-\frac{1}{2}\Delta_{z}\right)^{3}
-\frac{27}{8}\hbar\omega_{c}\Delta_{z}(2\alpha^{2}m)\right)
\end{eqnarray}
Equation (\ref{poly2}) has exactly one real solution if and only if the
discriminant
\begin{eqnarray}
D & = & \left(\frac{q}{2}\right)^{2}+\left(\frac{p}{3}\right)^{3}
=\left(\frac{2\left(\hbar\omega_{c}-\frac{1}{2}\Delta_{z}\right)}
{3\frac{e^{2}}{\kappa\ell}\sqrt{\frac{\pi}{8}}}\right)^{6}
\nonumber\\
&  & \cdot
\left(\left(1-\frac{\frac{27}{8}\hbar\omega_{c}\Delta_{z}(2\alpha^{2}m)}
{\left(\hbar\omega_{c}-\frac{1}{2}\Delta_{z}\right)^{3}}\right)^{2}-1\right)
\end{eqnarray}
is positive 
which is obviously the case for $\Delta_{z}<0$. Since the constant term
$q$ in Eq.~(\ref{poly2}) is also positive for $\Delta_{z}<0$ this single
real root has to be negative. It follows that Eq.~(\ref{poly2}) does not
have any positive solutions for $y$ if $\Delta_{z}<0$. The physical solutions
for $y$ have to lie in the interval $[r/3,1+r/3]$ and are ruled out if
$r>0$. The sign of $r$ is determined by the sign of
$(\hbar\omega_{c}+\Delta_{z})$. Therefore, to exclude the possibility
of physical solutions to Eq.~(\ref{poly2}) it is sufficient that the
cyclotron energy $\hbar\omega$ is larger in magnitude than the Zeeman
splitting $\Delta_{z}<0$, which is usually the case. In fact, both quantities 
are proportional to $|B|$, and their ratio depends 
just on the effective mass and the effective g-factor of the electrons.
In table \ref{table1} we have listed
parameter values for for conduction band electrons 
in typical III-V semiconductors which show that 
$\hbar\omega_{c}/|\Delta_{z}|$ is larger than unity 
for these materials. Thus, for the physically important case of
conduction band electrons in III-V semiconductors
no physical stationary point of the variational energy exists
except for the solution $c=b=0$, which corresponds to an energy maximum.
Therefore, the energy minimum must lie on the boundary defined by
$|b|^{2}+|c|^{2}=1$, and we step back to the restricted variational
state used in subsection \ref{case2} with $a=0$ (cf. (\ref{gensp})).
This finding also shows that the {\em ansatz} (\ref{ansatz1})
used in section \ref{case1} (involving a non-zero coefficient $a$) 
when applied to the case of subsection \ref{case2} cannot give 
a lower energy minimum than found by the {\em ansatz} (\ref{ansatz2}).


%
\begin{figure}
\centerline{\includegraphics[width=8cm]{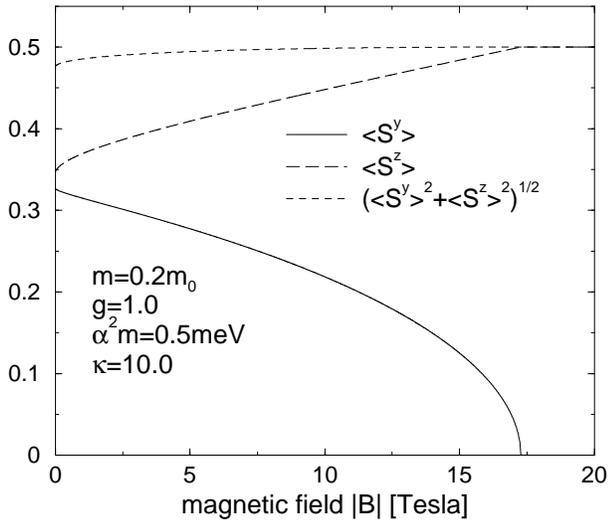}} 
\caption{Spin expectation values per particle (in units of $\hbar$)
as a function of $|B|$ for the same system parameters as in figure~\ref{fig1}.
The phase occuring in Eq.~(\protect{\ref{inplane}}) has been adjusted such that
$\langle s^{x}\rangle=0$ and $\langle s^{y}\rangle\geq 0$.
\label{fig3}}
\end{figure}
\begin{table}
\begin{tabular}{c|c|c|c|}
  &  $\frac{m}{m_{0}}$ & $g$ & $\frac{\hbar\omega_{c}}{|\Delta_{z}|}$ \\ 
\hline
GaAs & 0.067 & -0.44 & 67.8 \\
InAs & 0.023 & -15   & 5.7 \\
InSb & 0.014 & -51   & 2.8 \\
\end{tabular}
\caption{Effective masses (in units of the bare electron mass $m_{0}$) 
and g-factors for conduction band electrons 
in different III-V semiconductors. The last column shows that the ratio 
$\frac{\hbar\omega_{c}}{|\Delta_{z}|}$ is always larger than unity.
\label{table1}}
\end{table}


\begin{references}

\bibitem{carlosaddress}
Permanent address: Department of Physics and Informatics, University 
of S\~ao Paulo at S\~ao Carlos, 13560-970 S\~ao Carlos/SP, Brazil.

\bibitem{Wolf01}
S.~A. Wolf, D.~D. Awschalom, R.~A. Buhrman, J.~M. Daughton, S. von Molnar,
 M.~L. Roukes, A.~Y. Chtchelkanova, and D.~M. Treger, 
Science {\bf 294}, 1488 (2001).

\bibitem{Awschalom02}
{\em Semiconductor Spintronics and Quantum Computation}, eds.
D.~D. Awschalom, D. Loss, and N. Samarth, Springer, Berlin, 2002.

\bibitem{Rashba60}
E.~I. Rashba, Fiz. Tverd. Tela (Leningrad) {\bf 2}, 1224 (1960)
(Sov. Phys. Solid State {\bf 2}, 1109 (1960)); 
Y.~A. Bychkov and E.~I. Rashba, J. Phys. C {\bf 17}, 6039 (1984).

\bibitem{Dresselhaus55}
G. Dresselhaus, Phys. Rev. {\bf 100}, 580 (1955).  

\bibitem{Pikus88}
G.~E. Pikus, V.~A. Marushchak, and A.~N. Titkov, 
Sov. Phys. Semicond. {\bf 22}, 115 (1988).

\bibitem{Gerchikov92}
L.~G. Gerchikov and A.~V. Subashiev, Sov. Phys. Semicond. {\bf 26}, 73 (1992).

\bibitem{Datta90}
S. Datta and B. Das, Appl. Phys. Lett. {\bf 56}, 665 (1990). 

\bibitem{Das90}
B. Das, S. Datta, and R. Reifenberger, Phys. Rev. B {\bf 41}, 8278 (1990).

\bibitem{Falko92}
V.~I. Falko, Phys. Rev. B {\bf 46}, 4320 (1992);
Phys. Rev. Lett. {\bf 71}, 141 (1993).

\bibitem{Andrada97}
E.~A. de Andrada e Silva. G.~C. La Rocca, and F. Bassani,
Phys. Rev. B {\bf 55}, 16293 (1997).

\bibitem{Nitta97}
J. Nitta, T. Akazaki, H. Takayanagi, and T. Enoki, 
Phys. Rev. Lett. {\bf 78}, 1335 (1997).

\bibitem{Moroz99}
A.~V. Moroz and C.~H.~W. Barnes, Phys. Rev. B {\bf 60}, 14272 (1999);
A.~V. Moroz, K.~V. Samokhin,  and C.~H.~W. Barnes,
Phys. Rev. Lett. {\bf 84}, 4164 (2000).

\bibitem{Papadakis99}
S.~J. Papadakis, E.~P. De Poortere, H.~C. Manoharan, M. Shayegan, 
and R. Winkler, Science {\bf 283}, 2056 (1999).

\bibitem{Falko00}
V.~I. Falko and S.~V. Iordanskii, Phys. Rev. Lett. {\bf 84}, 127 (2000).

\bibitem{Winkler00}
R. Winkler, Phys. Rev. B {\bf 62}, 4245 (2000).

\bibitem{Matsuyama00}
T. Matsuyama, R. K\"ursten, G. Mei{\ss}ner, and U. Merkt, 
Phys. Rev. B {\bf 61}, 15588 (2000).

\bibitem{Grundler00}
D. Grundler, Phys. Rev. Lett. {\bf 84}, 6074 (2000).

\bibitem{Grundler01}
D. Grundler, Phys. Rev. Lett. {\bf 86}, 1058 (2001);
U. Z\"ulicke and C. Schroll, Phys. Rev. Lett. {\bf 88}, 029701 (2002);
D. Grundler, Phys. Rev. Lett. {\bf 88}, 029702 (2002).

\bibitem{Hausler01}
W. H\"ausler, Phys. Rev. B {\bf 63}, 121310 (2001).

\bibitem{Mireles01}
F. Mireles and G. Kirczenow, Phys. Rev. B {\bf 64}, 012426 (2001).

\bibitem{Aleiner01}
I.~L. Aleiner and V.~I. Falko, Phys. Rev. Lett. {\bf 87}, 256801 (2001).

\bibitem{Pareek01}
T.~P. Pareek and P. Bruno, Phys. Rev. B {\bf 65}, 241305(R) (2002).

\bibitem{Molenkamp01}
L.~W. Molenkamp, G. Schmidt, and G.~E.~W. Bauer,
Phys. Rev. B {\bf 64}, 121201 (2001).

\bibitem{Nitta02}
J. Nitta, T. Akazaki, and H. Takayanagi,  
Phys. Rev. Lett. {\bf 89}, 046801 (2002).

\bibitem{Matsuyama02}
T. Matsuyama, C.-M. Hu, D. Grundler, G. Meier, and U. Merkt, 
Phys. Rev. B {\bf 65}, 155322 (2002).

\bibitem{Governale02}
M. Governale and U. Z\"ulicke, Phys. Rev. B {\bf 66}, 073311 (2002).

\bibitem{Egues02}
J.~C. Egues, G. Burkard, and D. Loss, 
Phys. Rev. Lett. {\bf 89}, 176401 (2002).

\bibitem{Miller02}
J.~B. Miller, D.~M. Zumb\"uhl, C.~M. Marcus, Y.~B. Lyanda-Geller,
D. Goldhaber-Gordon, K. Campman, and A.~C. Gossard, cond-mat/0206375. 

\bibitem{Pinczuk97} 
For a review on experimental research see J.~P. Eisenstein
in {\it Perspectives in Quantum Hall Effects}, edited by S. Das Sarma and
A. Pinczuk, Wiley 1997; for a review on theoretical research see
S.~M. Girvin and A.~H. MacDonald, in the same volume.

\bibitem{Jungwirth01}
T. Jungwirth and A.~H. MacDonald, Phys. Rev. B {\bf 63}, 035305 (2001).

\bibitem{Girvin00}
S.~M. Girvin, Physics Today {\bf 53}, 39 (2000).

\bibitem{Spielman01}
I.~B. Spielman, J.~P. Eisenstein, L.~N. Pfeiffer, and K.~W. West,
Phys. Rev. Lett. {\bf 84}, 5808 (2000); 
{\it ibid} {\bf 87} 036803 (2001);
for a recent overview see also 
S.~M. Girvin, cond-mat/0202374, to appear in Proceedings of the Nobel 
Symposium on Quantum Coherence, Goteborg, Sweden, December, 2001 
(Physica Scripta).

\bibitem{Zheng97} 
L. Zheng, R.~J. Radtke, and S. Das Sarma, 
Phys. Rev. Lett. {\bf 78}, 2453 (1997).

\bibitem{MacDonald99} 
A.~H. MacDonald, R. Rajaraman, and T. Jungwirth,
Phys. Rev. B {\bf 60}, 8817 (1999).

\bibitem{Schliemann00}
J. Schliemann and  A. H. MacDonald, Phys. Rev. Lett. {\bf 84}, 4437 (2000).

\bibitem{DasSarma01}
S. Das Sarma and E. Demler,  Solid State Commun. {\bf 117}, 141 (2001).

\bibitem{Jungwirth98}
T. Jungwirth, S.~P. Shukla, L. Smrcka, M. Shayegan, and A.~H. MacDonald,
Phys. Rev. Lett. {\bf 81}, 2328 (1998).

\bibitem{Usher90}
A. Usher, R.~J. Nicholas, J.~J. Harris, and C.~T. Foxon, 
Phys. Rev. B {\bf 41}, 1129 (1990).

\bibitem{Maude96}
D.~K. Maude, M. Potemski, J.~C. Portal, M. Henini, L.~A. Eaves, G. Hill,
and M.~A. Pate, Phys. Rev. Lett. {\bf 77}, 4604 (1996).

\bibitem{Leadley97}
D.~R. Leadley, R.~J. Nicholas, D.~K. Maude, A.~N. Utjuzh, J.~C. Portal, 
J.~J. Harris, and C.~T. Foxon, Phys. Rev. Lett. {\bf 79}, 4246 (1997).

\bibitem{Zhitomirsky02}
V. Zhitomirsky, R. Chugtai, R.~J. Nicholas, and M. Henini,
Physica E {\bf 12}, 12 (2002).

\bibitem{note1}
A particle with with positive charge $q=|e|>0$ can formally be viewed as
a hole in a semiconductor quantum well. Note, however, that the spin orbit
interaction for holes in the valence band of most semiconductor materials 
is more complicated than for electrons in the conduction band whose
effective spin-orbit interaction is described by (\ref{soham}).
For a more complete theory of spin-orbit coupling in semiconductors
including holes see, e.g., Refs.~\cite{Pikus88,Gerchikov92}. 
In the present work, however, we shall concentrate on the effective
Hamiltonians (\ref{soham}) and discuss both cases given by the
sign of the product $gq$ of effective g-factor $g$ and charge $q$,
since for conductions band electrons $g$ can take both signs
\cite{Ivchenko97,Jiang01,Salis01}. 

\bibitem{Ivchenko97}
E.~L. Ivchenko, A.~A. Kiselev, and M. Wilander, 
Solid State Commun. {\bf 102}, 375 (1997).

\bibitem{Jiang01}
H. Jiang and E. Yablonovitch, Phys. Rev. B {\bf 64}, 041307(R) (2001).

\bibitem{Salis01}
G. Salis, Y. Kato, K. Ensslin, D.~C. Driscoll, A.~C. Gossard, 
and D.~D. Awschalom, Nature {\bf 101}, 619 (2001).

\bibitem{Recher00}
P. Recher, E.~V. Sukhorukov, and D. Loss, 
Phys. Rev. Lett. {\bf 85}, 1962 (2000).

\bibitem{Vandersypen02}
L.~M.~K. Vandersypen, R. Hanson, L.~H. Willems van Beveren, J.~M. Elzerman, 
J.~S. Greidanus, S. De Franceschi, and L.~P. Kouwenhoven, quant-ph/0207059.

\end{references}
\end{document}